\documentclass[12pt]{article}\usepackage[hyperfootnotes=false]{hyperref}
\usepackage{epsfig}
\usepackage{float}
\usepackage{empheq}
\usepackage{bbold}
\usepackage[utf8]{inputenc}
\usepackage{amsmath}
\usepackage{caption}
\usepackage{amsmath}
\usepackage{amssymb}
\usepackage{graphicx}

\newlength{\abstractwidth}
\renewcommand{\thefootnote}{\fnsymbol{footnote}}
\renewcommand{\thanks}[1]{\footnote{#1}} 
\newcommand{\starttext}{
\setcounter{footnote}{0}
\renewcommand{\thefootnote}{\arabic{footnote}}}

\newcommand{\be}{\begin{equation}}
\newcommand{\bea}{\begin{eqnarray}}
\newcommand{\eea}{\end{eqnarray}}
\newcommand{\beq}{\begin{equation}}
\newcommand{\ee}{\end{equation}}

	\newcommand*\widefbox[1]{\fbox{\hspace{2em}#1\hspace{2em}}}

	\def\dsp.{de Sitter space.}
	\def\eq{&=&}

	\def\la{\langle}
	\def\ra{\rangle}
	\def\simleq{\; \raise0.3ex\hbox{$<$\kern-0.75em
			\raise-1.1ex\hbox{$\sim$}}\; }
	\def\simgeq{\; \raise0.3ex\hbox{$>$\kern-0.75em
			\raise-1.1ex\hbox{$\sim$}}\; }
	
	\def\bi{\begin{itemize}}
		\def\ei{\end{itemize}}

	\def\dof{degrees of freedom }

	\def\CJ{{\cal{J}}}

	\def\CO{{\cal{O}}}

	\def\bx{{\bar{\chi}}}

	\def\bsub{ \begin{subequations}
			\begin{empheq}[box=\widefbox]{align}  }
			\def\esub{ \end{empheq}
	\end{subequations}}
	
	\def\1{\(  \mathbb{1} \)}

	\def\lf{\left(}
	\def\rg{\right)}

	\def\dk{${\rm DSSYK_{\infty}}$}
	
  \def\bx{\begin{equation} \boxed  }
  \def\L{\ell_{\mathrm{dS}}}
   \def\Ls{\ell_{\mathrm{string}}}
      \def\Lm{\ell_{\mathrm{micro}}}
      \def\R{r_{\mathrm{stretch}}}
	
	\makeatletter
	\g@addto@macro\normalsize{%
		\setlength\abovedisplayskip{10pt}
		\setlength\belowdisplayskip{20pt}
		\setlength\abovedisplayshortskip{10pt}
		\setlength\belowdisplayshortskip{20pt}
	}
	\makeatother
	
	\usepackage{color}

	\usepackage{ulem}
	\usepackage{authblk}
	\title{Infinite Temperature is Not So Infinite: \\
	 The Many Temperatures of de Sitter Space}

	\author[1]{Adel Rahman}
	\author[1,2]{ Leonard Susskind}
	\affil[1]{Stanford Institute for Theoretical Physics and Department of Physics\\ Stanford University, Stanford, CA 94305-4060, USA \vspace{1em}}
	\affil[2]{Google, Mountain View, CA}
	\date{}
	\newcommand{\inp} [1] {\left( #1 \right)}
	
	\newcommand{\insb} [1] {\left[ #1 \right]}
	
	\usepackage{upgreek}

\begin{document}
		
		\begin{titlepage}
			\maketitle
			
			\begin{abstract}
	
Several distinct concepts of temperature appear in the holographic description of de Sitter space. Conflating these has led to confusion and inconsistent claims. The double-scaled limit of SYK is a concrete model in which we can examine and explain these different concepts of temperature. This note began as an addendum to our paper ``Comments on a Paper by Narovlansky and Verlinde" but in the process of writing it we learned   new things---interesting in their own right---that we wish to report here.

			\end{abstract}

		\end{titlepage}
		
		\starttext 
		\setcounter{footnote}{0}
		
 \tableofcontents
	
\section{Introduction}

\quad This paper began as an addendum to our recent  paper ``Comments on a Paper by Narovlansky and Verlinde" \cite{Rahman:2023pgt} which was a response to the paper \cite{Narovlansky:2023lfz}. In  writing it we discovered  interesting and surprising new things about the various concepts of temperature that appear in the holographic description of de Sitter space. We will illustrate them here  using the  \dk--de Sitter duality conjectured  in \cite{Susskind:2021esx,Susskind:2022bia,Rahman:2022jsf,Susskind:2023hnj,Verlinde}.
	
That there is more than one concept of temperature in the holographic description of de Sitter space
%
became apparent when it was argued that the entanglement spectrum of the de Sitter static patch is flat \cite{Banks:2005bm,Banks:2006rx,Fischler,Dong:2018cuv,Susskind:2021omt,Chandrasekaran:2022cip}. That fact requires the ``Boltzmann temperature" 
appearing in the thermal density matrix to be infinite in \it cosmic \rm units\footnote{The Boltzmann temperature being infinite in cosmic units allows for it to be finite in string units. In that case the maximal mixing condition is slightly violated by $\sim$ 1 bit. Fractionally, the violation is of order $1/N$, see Section \ref{corrections}.} (i.e. units adapted to the de Sitter scale $\ell_{\mathrm{dS}}$)---hence the $\infty$ in DSSYK$_{\infty}$. However, the Hawking temperature in cosmic units is finite,  implying that there are at least two temperatures to keep track of.

Two other notions of temperature have appeared in various contexts. The first was the  ``Tomperature" \cite{Lin:2022nss}, 
whose definition we will review in Section \ref{Tomp} below. The second was the ``fake-disc" temperature introduced in \cite{Lin:2023trc} which we will often call the ``cord" temperature 
for reasons to be explained below. Although 
Tomperature and fake disk/cord temperature were first encountered in the context of the double-scaled SYK model, they are likely to be far more general features of de Sitter holography. As we will explain below, the Tomperature 
is simply another avatar of the coordinate Hawking temperature (which will agree with the physical Hawking temperature experienced at the pode) whereas the fake disk/cord temperature is the physical Hawking temperature experienced at the stretched horizon.

To complicate matters,  there are at least two relevant systems of units---\textit{cosmic} and \textit{string}---which are separated from one another by scale transformations that diverge in the semiclassical limit\footnote{We use the term \it semiclassical limit\rm \ in the weak sense described in the appendix of \cite{Rahman:2023pgt}. In the weak semiclassical limit gravity behaves semiclassically at large scales while matter remains fully quantum mechanical; by contrast, in the strong semiclassical limit \textit{all} quantum fluctuations tend to zero.}; this is, of course, the ``separation of scales" which is known to occur in the semiclassical limit of de Sitter space \cite{Susskind:2022bia}.  
Cosmic units are adapted to the de Sitter length $\ell_{\mathrm{dS}}$ while string units are adapted to the string length $\ell_{\mathrm{string}} $. For example, a quantity $L$ with units of length would have numerical value $L/\ell_{\mathrm{dS}}$ in cosmic units and numerical value $L/\ell_{\mathrm{s}}$ in string units. In the semiclassical limit, the ratio of these scales diverges 
\begin{equation}
	\frac{\ell_{\mathrm{dS}}}{\ell_{\mathrm{string}}} \ \underset{\text{semiclassical limit}}{\longrightarrow} \ \infty
\end{equation}
which is the origin of the separation of scales: a quantity with units of length which is finite in string units will tend to zero in cosmic units, while a quantity with units of length which is finite in cosmic units will tend to infinity in string units\footnote{In cosmic units the  curvature of de Sitter space, the energies of  Hawking quanta, and the frequency of quasinormal modes all remain finite in the semiclassical limit;  while the masses of elementary particles, excitation energies of strings, et cetera become infinite (or scale with $N$). By contrast, in string units the radius of curvature of de Sitter space, wavelengths of Hawking quanta, and periods of  quasinormal modes diverge in the semiclassical limit; while the compton wavelengths of particles, periods of string oscillations, et cetera remain finite.}.

Below is a chart (Fig. \ref{table}) to help navigate through the various temperatures and unit systems that will appear in this paper
\begin{figure}[H]
	\begin{center}
		{\renewcommand{\arraystretch}{1.15}
			\large
			\begin{tabular}{c|c|c|}
				& Cosmic Units ($\uptau$) & String Units ($T$)\\
				\hline
				\begin{tabular}{c}
					\ \\
					Boltzmann Temperature \\
					\
				\end{tabular} & $\uptau_B \ \sim \ \infty$ & $T_B \ \sim \ \dfrac{\infty}{\infty}$ \\
				\hline
				\begin{tabular}{c}
					\ \\
					Hawking Temperature \\ $\sim \ $ Tomperature
					\\
					\ 
				\end{tabular}  & 
				$\uptau_H = 2\mathcal{J}_0$ & $T_H \ \sim \ \dfrac{\mathcal{J}_0}{p}$\\
				\hline
				\begin{tabular}{c}
					\ \\
					cord Temperature \\
					$\equiv$ ``Fake Disk" Temperature\\
					\
				\end{tabular}
				& $\uptau_{\mathrm{cord}} \ \sim \ p\mathcal{J}_0$ & $T_{\mathrm{cord}} = \dfrac{\mathcal{J}_0}{\pi}$\\
				\hline
			\end{tabular}
		}
		\caption{The various types of temperature which appear in the analysis of DSSYK$_{\infty}$ along with their values in the two major units systems (cosmic and string).}
		\label{table}
	\end{center}
\end{figure}
Here $p$ is the k-locality parameter of the SYK model (which tends to infinity in the double-scaled limit) and $\mathcal{J}_0$ is a number characterizing the variance of the random couplings in the DSSYK model, described in \eqref{variance1} below\footnote{$\mathcal{J}_0$ agrees with the numerical value of the parameter $\mathcal{J}$ that appears in the usual SYK literature (see e.g. \cite{Maldacena:2016hyu}) in which the primary focus is on string units.}. \textbf{We have also introduced the following notation:} In this paper we will denote (the numerical value of) temperatures in cosmic units by Greek $\uptau$'s and temperatures in string units by uppercase Latin $T$'s. This is meant to promote agreement with the usual literature, in which the primary focus is on string units. More generally, as in \cite{Rahman:2023pgt}, we will also use the notation:
	\be  
	[A]_{units}
	\ee
to represent the numerical value of the dimensionful quantity $A$ in the unit system $units.$ For example, we have 
$$\uptau_B \equiv [\text{Boltzmann Temperature}]_{\mathrm{cosmic}}$$
and 
$$T_B \equiv [\text{Boltzmann Temperature}]_{\mathrm{string}}$$
et cetera. We will denote the Hawking temperature by $\uptau_H$/$T_H$ and the cord temperature by $\uptau_{\mathrm{cord}}$/$T_{\mathrm{cord}}$.

We can translate the formulas in the table (figure \ref{table}) into bulk expressions by making use of the dictionary item
\begin{equation}
	2\mathcal{J}_0 = \frac{1}{2\pi\ell_{\mathrm{dS}}}
\end{equation}
The other important item from the dictionary will be
\begin{equation}
	\frac{\ell_{\mathrm{dS}}}{\ell_{\mathrm{string}} } \ \sim \ p 
	\label{dSs}
\end{equation}
see \cite{Susskind:2022bia,Rahman:2023pgt} for more details.
Here we have adapted the notation 
\begin{equation}
	A \ \sim \ B
\end{equation}
from \cite{Rahman:2023pgt} which is taken to mean that  the quantity $A$ scales parameterically as the quantity $B$ in the large $N$/(weak) semiclassical limit.

The chart in figure \ref{table} may be expressed in equation form as
\begin{equation}
	\begin{aligned}
		\uptau_B  \ &\sim \    \infty       \cr \cr
		T_B   \ &\sim \    \infty/\infty      \cr \cr
		\uptau_H  \ &= \    2\CJ_0       \cr \cr
		T_H \ &\sim \    \CJ_0/p      \cr \cr
		\uptau_{\mathrm{cord}}  \ &\sim \    p\CJ_0          \cr \cr
		T_{\mathrm{cord}}  \ &= \    \CJ_0/\pi        
	\end{aligned}
\end{equation}
Notice that temperatures in cosmic and string units differ by a factor of $p$, which tends to infinity in the double-scaled limit. This represents the relationship \eqref{dSs}, i.e. that the ratio of cosmic and string length scales is of order $p.$ The ambiguous nature of the Boltzmann temperature will be explained as we proceed.

Time is a dimensionful quantity which transforms inversely to energy. For example time intervals in string and cosmic units are related by,
\be 
[\Delta t]_{\mathrm{string}} \ \sim \ p\cdot[\Delta t]_{\mathrm{cosmic}}
\label{ts=ptc}
\ee
(e.g. a time interval which is $O(1)$ in cosmic units will be extremely long, $\sim O(p)$ in string units). To simplify the notation for time, we will find it helpful to define
\bea 
t_s &\equiv& [t]_{\mathrm{string}}  \cr \cr
t_c &\equiv& [t]_{\mathrm{cosmic}}
\label{t-convs}
\eea

In what follows we will assume all of the conventions and notations of \cite{Rahman:2023pgt} with one exception: Equations will not be boxed to   distinguish correct equations from incorrect ones, except in the quotation below. Other than that  we intend to only write correct equations here. We begin by quoting from  \cite{Rahman:2023pgt}: 
	
	\it
There are three distinct concepts of temperature that appear in the holographic formulation of de Sitter space. These seem to be different and not related by just a change of units. The first is the ``Boltzmann temperature" $T_B$ which is the temperature parameter that appears in the thermal density matrix,
\be \boxed{
\rho = \frac{1}{Z}\,\exp{\inp{-H/T_B}}}
\label{density}
\ee
What we know about $T_B$ is that it is infinite in cosmic units
\be \boxed{ 
[T_B]_{\mathrm{cosmic}} = \infty}
\label{Tb=infty}
\ee
...
Indeed the $\infty$ in DSSYK$_{\infty}$ is meant to refer to the value of $T_B$.

We cannot conclude from this that the Boltzmann temperature is also infinite in string units since the ratio of the cosmic scale to the string scale goes itself to $\infty$ in the double-scaled limit.
Indeed there are reasons to believe that \eqref{Tb=infty} should be refined to read,
\be
[T_B]_{\mathrm{cosmic}} \ \sim \ p\CJ_0 \ \to \ \infty  \quad \text{(in double-scaled limit)}
\label{Tb=J0p}
\ee
This would change nothing in the analysis of cosmic-scale phenomena but can affect $1/N$ corrections to string-scale phenomena. Changing  to string units in \eqref{Tb=J0p}  gives,
\be 
[T_{B}]_{\mathrm{string}} \ \sim \ \CJ_0 \qquad \text{(speculative)}
\label{Tbstring}
\ee
Here we see an interesting point: infinite temperature in cosmic units does not necessarily mean infinite temperature in string units. The infinity in DSSYK$_{\infty}$ should always be interpreted as infinite temperature in cosmic units.
For now this is simply an aside, but we will return to this point in a future publication. \rm \\
	
	This note is the ``future publication" referred to above.	
	
	\section{The Boltzmann Temperature and Corrections to Entropy }\label{corrections}
	
\quad We will consider 
\begin{equation}
	T_B \equiv [\text{Boltzmann Temperature}]_{\mathrm{string}}
	\label{tauB}
\end{equation}
to be a free parameter of the double-scaled SYK theory. For reasons that we will explain in section \ref{S:fake} $T_B$ should be somewhat larger than $\mathcal{J}_0$ in order to be in the de Sitter regime.

The value of  $T_B$ has nontrival effects for finite $N$. For any nonzero value of $T_B$ the Boltzmann temperature in cosmic units is infinite in the large $N$ limit
\be 
\uptau_B \ \sim \ p\,T_B \ \to \  \infty.
\ee
but only in the limit $T_B\to \infty$ is it infinite for finite $N$.
At finite $N$, there seems to be two possible prescriptions: One prescription would be to set $T_B=\infty$;  another would be to let $T_B$ be finite. These two prescriptions would lead to the same infinite $N$ results but would differ when considering $1/N$ corrections. In this section we will illustrate this point by computing $1/N$ corrections to the 
entropy. This ambiguity is the reason the Boltzmann temperature in string units was listed as ``$\infty/\infty$" in figure \ref{table} above.
		
Let us calculate the entropy of DSSYK at high but not infinite temperature
$$0 < \beta_B \ll 1.$$
where we have defined
\begin{equation}
	\beta_B \equiv \frac{1}{T_B}
\end{equation}
We begin by expanding the partition function in powers of the inverse Boltzmann temperature $\beta_B$:
\begin{align}
	Z 
	&= \mathrm{Tr}\, e^{-\beta_B H} \nonumber\\[0.25em]
	&= \mathrm{Tr}\, \lf  1 -\beta_B H +\frac{1}{2}\,\beta_B^2 H^2+ \cdot\cdot\cdot\rg
\end{align}

Here we are working in string units so $H \equiv [H]_{\mathrm{string}}$. The first two terms are trivial:
	\bea 
\mathrm{Tr}\, 1 \eq 2^{N/2} \cr \cr
\mathrm{Tr}\, H \eq 0  .
\label{traces}
\eea
The third term,  $\mathrm{Tr}\, H^2$, is easily calculated. The Hamiltonian is given in string units\footnote{Recall that what we are calling ``string units" correspond to the ``usual" conventions for SYK that are used in most of the usual literature, see e.g. \cite{Maldacena:2016hyu}.}, by
\begin{equation}
			[H]_{\mathrm{string}} \ = \sum_{i_1 < i_2 < \,\dots\, < i_p}  [J_{i_1i_2\,\dots\, i_p}]_{\mathrm{string}}\,\psi_{i_1}\psi_{i_2}\dots\psi_{i_p}
			\label{Hx}
\end{equation}
with random couplings drawn from a Gaussian ensemble of variance
\be 
[\la  J^2 \ra]_{\mathrm{string}} =   \frac{N}{p^2}  \frac{ \CJ_0^2}{ {N \choose p} }     
=   \frac{2}{\lambda} \frac{ \CJ_0^2}{ {N \choose p} }  .
\label{variance1}
\ee
One easily finds that
\be  
[\mathrm{Tr}\, H^2]_{\mathrm{string}} = 
2^{N/2} \lf \frac{2\CJ_0^2}{\lambda} \rg.
\ee
To order $\beta_B^2$, we then have that
\be 
Z = 2^{N/2}  \inp{1 + \frac{\mathcal{J}_0^2}{\lambda T_B^2} + \cdot\cdot\cdot}
\ee 
The free energy 
\be 
F = T_B \log{Z}
\ee
is then given by
\be 
[F]_{\mathrm{string}} = \frac{\log(2)}{2}\,N\,T_B + \frac{\mathcal{J}_0^2}{\lambda T_B} + \dots
\ee
Using the thermodynamic relation $S = \mathrm{d}F/\mathrm{d}T_B,$ we find that
\begin{equation}
	S =  \frac{\log(2)}{2}\,N -	\frac{1}{\lambda}\inp{\frac{\mathcal{J}_0}{T_B}}^2
\end{equation}
So we see that slightly lowering the temperature leads to a decrease
\begin{equation}
	\Delta S = -\frac{1}{\lambda}\inp{\frac{\mathcal{J}_0}{T_B}}^2
	\label{SfrGR}
\end{equation}
relative to its infinite temperature value $\frac{\log(2)}{2}\,N$.
	
We see from this example that choosing $T_B$ finite (as opposed to infinite) affects $1/N$ corrections to the semi-classical limit. In this case the correction is of order a single bit for $T_B\sim \CJ_0$ (and, of course, for $\lambda \sim O(1)$, as is part of our definition of the double-scaled limit). The correction to the maximal mixing of the density matrix will then also be very small. 
The fractional correction $\Delta S/S$ will be of order $1/N$,  so that the \textit{relative} size of such corrections will be extremely small for $N \gg 1.$

\section{ Tomperature and Hawking Temperature}
\label{Tomp}
\quad The \textit{Tomperature} 
was introduced by Lin and Susskind in the de Sitter context in \cite{Lin:2022nss}. The notion of Tomperature  is distinct from that of the Boltzmann temperature although both are defined through the familiar first law,
\be 
\Delta E = T \Delta S.
\label{frstlaw}
\ee
They differ because the meaning of the incremental change $\Delta S$ is different in the two cases. In the case of Boltzmann temperature the incremental change in entropy  refers to a change in which the number of degrees of freedom is held fixed while a change is made in the energy (as defined by the SYK Hamiltonian). By contrast the Tomperature is defined by removing or freezing\footnote{By ``freezing" a qubit we mean projecting it onto some pure state and then holding it fixed.} a qubit (Fermion pair) while keeping fixed  the couplings of all other Fermions. This mimics what happens when a Hawking quantum is emitted into the bulk of the static patch, decoupling from the stretched horizon. In other words, we expect that the notion of Tomperature holographically encodes the bulk notion of Hawking temperature:
\begin{equation}
	\text{Tomperature $\ \sim \ $ Hawking Temperature}
	\label{tompHawk}
\end{equation}
Here by ``Hawking Temperature" we mean the coordinate temperature in the usual static patch coordinates \eqref{metric} or, what is equivalent, the physical temperature experienced at the pode (center of the static patch). In \eqref{tompHawk} we use ``$\sim$" rather than ``$=$" since the tomperature as strictly defined above might be off from the Hawking temperature by some $O(1)$ factor (e.g. a Hawking quanta might not precisely correspond to a single qubit). The main point is that both notions of temperature capture the same qualitative physics and exist at the cosmic scale, i.e. are finite in cosmic units. We will recover the precise holographic value of the Hawking temperature---including the overall $O(1)$ factor---by studying single-Fermion correlators in Section \ref{section Hawking} below.

Freezing a single qubit  changes the entropy of the DSSYK$_{\infty}$ system/holographic degrees of freedom by $\Delta S \sim-1$. In \cite{Lin:2022nss} Lin and Susskind---working in cosmic units---showed that the corresponding energy change is given by $[\Delta E]_{\mathrm{cosmic}} = 2\CJ_0.$ Thus the Tomperature  is given, in cosmic units, by
\be 
\uptau_H \ \sim \ \mathcal{J}_0
\label{tomp eq}
\ee

The Tomperature was defined on the DSSYK$_{\infty}$ side of the duality but for reasons we explained it has a bulk interpretation as the (coordinate/pode) Hawking temperature. This will provide an important bridge between the two sides of the duality.

\section{Chords, Cords, and Strings}
\quad Chord operators \cite{Berkooz:2018jqr}  are multi-Fermion operators of Fermion-weight $p\Delta$ where $\Delta$ is the so-called dimension of the chord. $\Delta$ is assumed to be parameterically of order unity. There are two kinds of chords: Hamiltonian chords and matter chords. From here on we will refer to  matter chords as \it cords, \rm partly  to distinguish them from Hamiltonian chords and partly to emphasize their similarity to strings. The energy scale of cords is the same as the string scale, 
$\ell_{\mathrm{string}} ^{-1}$ ($ = M_{\mathrm{string}}$ in the notation of \cite{Rahman:2023pgt}).

Generic cord operators 
\be 
\mathcal{O}_{\mathrm{chord}} =\sum_{i_1<i_2...<i_{\Delta p}} K_{i_1 i_2 ...i_{\Delta p}} \,\psi_{i_1}\psi_{i_2}...\psi_{i_{\Delta p}}
\label{mcord}
\ee
are defined by a dimension $\Delta$ and a set of random couplings $K_{i_1 i_2 ...i_{\Delta p}}$  drawn from a Gaussian random ensemble with variance \cite{Berkooz:2018jqr},
\be 
\la K^2 \ra \ \sim  \ {N \choose \Delta p}^{-1}.
\label{varM}
\ee

In the semiclassical limit of de Sitter space there is a broad range of energies/masses $M$ for which both the curvature of de Sitter space and gravitational backreaction can be ignored \cite{Susskind:2023hnj,Rahman:2023pgt}; within that energy range, phenomena can be treated as if in flat spacetime. This flat space region is centered on the micro-scale
\begin{equation}
	[M_{\mathrm{micro}}]_{\mathrm{cosmic}} \ \sim \ \sqrt{N}  \CJ_0 .
\end{equation}
In the double-scaled limit $p \sim \sqrt{N}$ the string/cord scale
\begin{equation}
	[M_{\mathrm{string}}]_{\mathrm{cosmic}} \ \sim \ p \CJ_0
\end{equation}is deep  in the flat space region: Indeed the ratio of string-scale to micro-scale is parametrically order unity in the double-scaled limit,
\be 
\frac{M_{\mathrm{string}}}{M_{\mathrm{micro}}} \ \sim \ \sqrt{\lambda}.
\ee
The physics of cords therefore effectively takes place in flat space. We expect that this flat space cord-theory is analogous to flat space string-theory. 

Why then don't we study cord-theory in flat space, for example in the conventional light-cone frame? We should, but the problem is that cord physics is presented to us in a very awkward holographic format; namely in the context of static patch holography\footnote{This could be either a static patch of de Sitter space, or, for low temperatures, the more familiar case of the static patch of an AdS$_2$ black hole. The main point is that in either case, there is a horizon for which the Hamiltonian is locally a boost generator.}. The flat space limit of the static patch is the one-sided Rindler wedge. Cord physics may have a simple form in flat space 
Cartesian or 
light-cone coordinates but from our current understanding of DSSYK$_{\infty}$ we only know this theory in the unfamiliar setting of Rindler space. The bridge from Rindler space to the light-cone frame is unknown but  such a bridge must exist. Nevertheless, we do know some things: Among them is the fact that almost everything is ``confined"   \cite{Susskind:2023hnj}.

\subsection{Almost Everything is Confined} \label{sec confine}
\quad We expect that the multi-Fermion operators which create string-like cords which are able to escape the near horizon region and propagate deep into the bulk are very special ``singlet" operators. All other cord operators create excitations which are ``confined" to the near-horizon region \cite{Susskind:2023hnj}. Generic cords \eqref{mcord} will have projections onto the singlets but these projections will be very small (of order $1/N$ to some power). The usual process of ensemble-averaging will be overwhelmed by the non-singlets and miss the exceptional singlets.

In  \cite{Susskind:2023hnj} it was explained  that apart from the tiny 
number of singlets, generic operators  \eqref{mcord} create collections of \it unbound \rm Fermions\footnote{Unbound to each other but bound to the stretched horizon.  The term confined in the present context refers to whether an object is trapped or confined to the stretched horizon region and not whether it forms strong bonds with other similar objects. See  \cite{Susskind:2023hnj} for more details.} which are confined to the stretched horizon region.  They do not propagate into the bulk of the static patch. They are unbound because they live in a hot plasma-like region---the stretched horizon---where the thermal energy is enough to dissociate them into their constituents\footnote{\label{factorize} More precisely, cord correlators factorize into products of single Fermion correlators to leading order in the cord coupling $\lambda$. Our point of view is that these subleading non-factorizing terms come from normal interactions which are not strong enough to bind cords into long-lived composite particles. In fact, it seems likely that these $O(\lambda)$ interactions are simply ordinary gravitational interactions, since for generic cords of mass $M_{\mathrm{cord}} \sim p\mathcal{J}_0$, we have using the dictionary in \cite{Rahman:2023pgt} that 
\begin{equation}
	\frac{G M_{\mathrm{cord}}^2}{\mathrm{distance}} \ \sim \ \frac{p^2}{N} \ \sim \ \lambda
\end{equation}
} \cite{Susskind:2023hnj}.

As previously mentioned, it was further conjectured in \cite{Susskind:2023hnj} that special $O(N)$ (or in the case of complex SYK, $SU(N)$) singlet operators  can escape the stretched horizon region and  propagate into the bulk. If all Fermions and cords could escape there would be far too many independent \dof \ propagating into the bulk, but
singlet operators are very rare in the space of all cord operators\footnote{While this type of confinement is not the subject of this paper it is crucial to the validity 
of the interpretation of DSSYK$_{\infty}$ as a theory of de Sitter space.}. Conventional ensemble averaging would entirely miss them.

\subsection{Singlets and the Flat Space Limit: A Challenge}
\quad Singlet cord operators do exist. A subset is defined by\footnote{Such operators were studied by Gross and Rosenhaus in \cite{Gross:2017aos}.}
\be 
\CO_n = \sum_i\psi_i\,\frac{\mathrm{d}^n \psi_i}{\mathrm{d}t^n}.
\label{singcord}
\ee
Naively the $\mathcal{O}_n$ look like two-Fermion operators but there are hidden actions of the Hamiltonian in taking the time-derivatives. In fact they form a tower of  operators of dimension $\Delta =n.$ We expect that these singlets  behave like strings propagating in the bulk. In contrast to the singlets,  
 generic cords are simply part of the stretched horizon ``soup". 

Singlet cords are expected to be capable of probing the bulk geometry but isolating them is more difficult than studying generic cords. The usual method of ensemble averaging will lose this signal because singlets are very sparse in the space of cords. But our conjecture is that these singlets are what survive in the flat (Rindler) space limit far from the horizon; and that their properties are similar to strings. In particular in the $\lambda \to 0$ limit (or, more precisely, in the $N\to \infty$ limit with fixed $p$ followed by the $p \to \infty$ limit), we conjecture that cords behave similarly to  strings  in the limit of vanishing string coupling. In other words we conjecture that there is a theory  of  ``free cords" analogous to free string theory (in the limit of vanishing string coupling).

It should be possible---at least in principle---to formulate free cord theory  in more conventional coordinates, such as light-cone coordinates, in a manner similar to the formulation of BFSS theory. The carriers of longitudinal momentum---D0-branes---would be replaced by the fundamental Fermions, and the light-cone Hamiltonian would be drawn from a random ensemble. We do not have a detailed proposal but raise the possibility as a challenge.

\subsection{Back to Generic Cords}
\quad \ Returning to  generic cords, their properties  suggest that they live in a hot environment in which they ``melt" into constituent fundamental Fermions. This requires a temperature of order unity in string units; in other words this requires a local temperature,
\be 
T_{\mathrm{cord}} \sim \CJ_0
\ee 

Three facts support this view of generic cords. 
\begin{itemize}
\item  The first is that cord correlation functions factorize into products of Fermion correlations \cite{Lin:2023trc} (at least to leading order in $\lambda$, see footnote \ref{factorize}). This is the basis for the claim that generic cords trivially behave like collections of non-interactiong Fermions\footnote{We expect this to be true for the high temperature limit of DSSYK. We make no claim for the more commonly studied low temperature limit.}. Similar things would be true for gauge theory quanta (quarks and gluons) in a hot QCD plasma.

\item The second fact, which we will explain in Section \ref{section Hawking} below, is that the generic cord correlation functions exponentially decay without oscillating (when time is measured in cosmic units). This type of decay is also characteristic of particles in a hot plasma; if the plasma is hot and dense enough the correlations will be overdamped.

\item 
Finally, cord correlation functions are periodic in imaginary time with period $\sim \mathcal{J}_0^{-1}$ \cite{Lin:2023trc}. For example, at\footnote{\label{complicated} For finite $\uptau_B$, the Euclidean two-point function is more complicated, given by \cite{Maldacena:2016hyu,Lin:2023trc}
\be
\la \, \mathcal{O}(\tau_s)\,\mathcal{O}(0) \,\ra \sim \inp{\frac{\cos\frac{\pi v}{2}}{\cos\insb{\frac{\pi v}{2}\inp{1-2T_B\tau_s}}}}^{\Delta}
\ee
In either case, we are working with simple $\lambda \to 0$ expressions for the cord two-point function in order to illustrate the basic point. It is known that this periodicity survives at least to the leading $O(\lambda)$ correction to the correlation function. We are very grateful to H. Lin for discussions on this point.} 
$T_B = \infty$, the Euclidean continuation
\be 
\la\,\mathcal{O}(\tau_s)\,\mathcal{O}(0)\,\ra \sim \lf \frac{1}{\cos^2{\CJ_0 \tau_s}} \rg^{\Delta}
\label{crdcorr}
\ee
of the cord two-point function is periodic with period $\pi/\mathcal{J}_0$ (here $\tau_s = \mathrm{i}t_s$ and we are using the conventions of \eqref{t-convs}), leading to an interpretation as a thermal two-point function with temperature \cite{Lin:2023trc}
\be 
T_{\mathrm{cord}} = \frac{\mathcal{J}_0}{\pi}
\label{Tcord}
\ee

\quad For finite $T_B$, the expression for the two-point function and the cord temperature become more complicated (see e.g. footnote \ref{complicated}), but---as we will show in the next section---for order-one (string-unit) temperatures $T_B \gtrsim \mathcal{J}_0$, the cord temperature rapidly asymptotes to its infinite temperature value 
\eqref{Tcord}. 

\quad Away from zero temperature, the cord temperature is distinct from and strictly smaller than the Boltzmann temperature, differing by a $T_B$-dependent factor
$0 < v(T_B) \leq 1$ to be explained in the next section: 
\begin{equation}
	T_{\mathrm{cord}} = v\,T_B
\end{equation}
\end{itemize}

\section{Real and Fake Disks}\label{S:fake}
\quad We now come to a central point of this paper  involving a loose end that we have yet to tie up, namely: how does $T_{\mathrm{cord}}$ depend on $T_B$?

The Boltzmann temperature and the cord temperature each have a Euclidean thermal circle  associated with them. When filled-in to form discs they are called the ``real disc" and the ``fake disc" respectively \cite{Lin:2023trc}. Tying this loose end is the same as determining the ratio of the sizes of the real and fake discs.

That the temperature experienced by generic cords should be smaller than the Boltzmann temperature was already pointed out earlier by Lin and Stanford \cite{Lin:2023trc}. In that context, they make use of a parameterization $v(T_B)$ of the Boltzmann temperature defined by the equation (see e.g. \cite{Maldacena:2016hyu} or \cite{Lin:2023trc})
\begin{equation}
	\frac{\pi v}{\mathcal{J}_0}\,T_B = \cos\frac{\pi v}{2}
	\label{mystery0}
\end{equation}
where we remind the reader that we are working in string units. The parameter $v$ runs from $1$ (at $T_B = 0$) down to $0$ (at $T_B = \infty$). A central result of \cite{Lin:2023trc} was that the temperature experienced by generic cords is not the Boltzmann temperature, but rather the cord temperature $T_{\mathrm{cord}} = vT_B$  giving
\begin{equation}
	v = 
	\frac{T_{\mathrm{cord}}}{T_{B}}  
	\label{v}
\end{equation}
(see also the arguments above).

Either equation \eqref{mystery0} or \eqref{v} has, by itself, no content, but combing the two equations gives a relation between $T_B$ and $T_{\mathrm{cord}}$, namely
\be  
\frac{\pi}{\CJ_0}\,T_{\mathrm{cord}} = \cos\inp{\frac{\pi}{2} \frac{T_{\mathrm{cord}}}{T_B}}
\label{mystery}
\ee

\begin{figure}[H]
		\begin{center}
			\includegraphics[scale=.5]{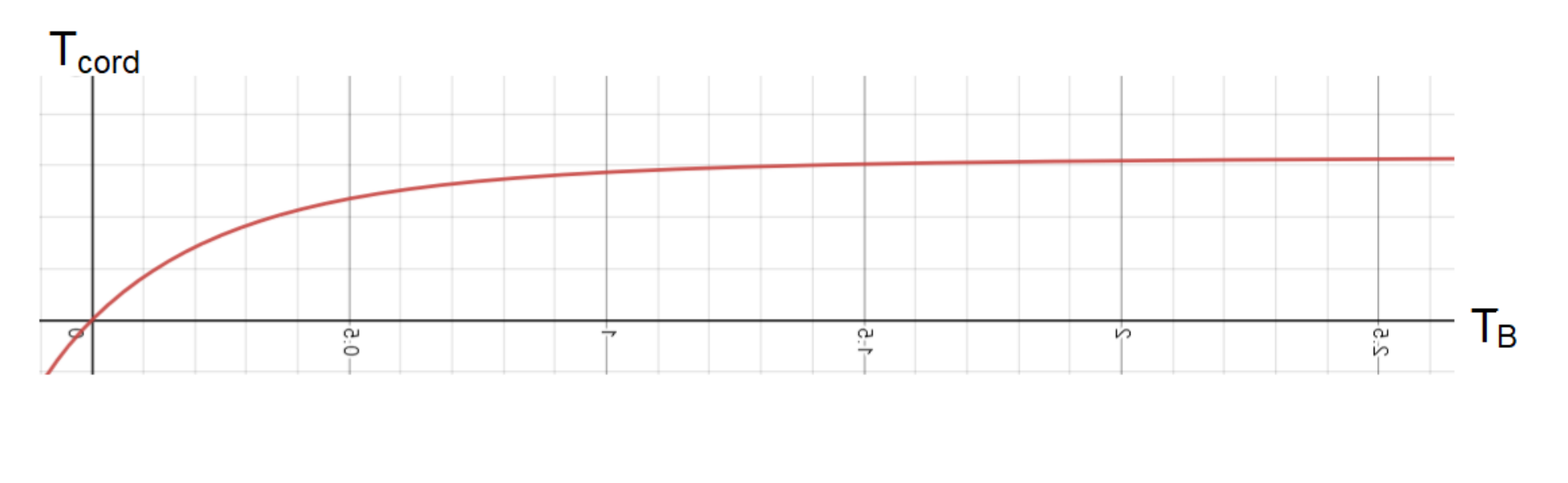}
			\caption{The relationship between $\uptau_{\mathrm{cord}}$ and $\uptau_B$ measured in multiples of $\mathcal{J}_0$.}
			\label{myst}
		\end{center}
	\end{figure}

In figure \ref{myst} the relation \eqref{mystery} is plotted. 

There are three regions: In the first, $T_B \lesssim .5\mathcal{J}_0$, the relation is  approximately linear. This is the AdS region. This then gives way to a transition region $.5\mathcal{J}_0 \lesssim T_B \lesssim \mathcal{J}_0,$ and then to an infinite plateau $T_B \gtrsim \mathcal{J}_0$ where $T_{\mathrm{cord}}$ is close to, but slightly less than its asymptotic value
\begin{equation}
	T_{\mathrm{cord}} \ \to \ \frac{\mathcal{J}_0}{\pi}
\end{equation}
This plateau is the de Sitter region. 

In the plateau region equation \eqref{SfrGR} may be written in the approximate form
\be 
\Delta S \ \sim \  -\frac{1}{\lambda} \lf  \frac{T_{\mathrm{cord}} }{T_B}   \rg^2
\label{SfrGRR}
	\ee
Since we always have that $T_{\mathrm{cord}}/{T_B} = v \sim O(1)$, we emphasize again that the correction to the entropy (relative to its value at $T_B = \infty$) is always roughly a single bit in the double-scaled limit $\lambda = \mathrm{finite}$.

\section{Correlation Functions and Hawking Temperature} \label{section Hawking}
\quad We saw in section \ref{Tomp} above that the Hawking temperature is a cosmic-scale object, scaling like 
\begin{equation}
	\uptau_{H} \ \sim \ \mathcal{J}_0
\end{equation}
In this section we will study two types of correlation functions---cord and single Fermion---and show that the latter have a characteristic dissipation time at the cosmic scale. We conjecture that this timescale is the inverse of the bulk Hawking temperature, providing the precise numerical factor in the holographic definition of $\uptau_H$. Validating this conjecture requires more thought, and we will return to this issue in future works.

These two types of correlation function---cord and single Fermion---naturally demonstrate the separation of scales. The former describe string-scale physics while the latter describe cosmic-scale physics.

Let's consider the Fermion two-point function $G$ with time measured in string units. For simplicity, we will work in this section using $T_B = \infty$ expressions, since we are only interested in the de Sitter region and have shown in the previous section that this is a good approximation for temperatures $T_B \gtrsim \mathcal{J}_0$. Using the notation of equation \eqref{t-convs}, we have
\be  
G(t_s) =\la \psi(t_s)\,\psi(0)\ra.
\label{G}
\ee
We begin with the cord 2-point function\footnote{Again, as previously explained in footnote \ref{complicated} above, we are working with simple $\lambda \to 0$ expressions for the two-point functions to illustrate basic points without getting lost in the mathematical weeds.}. For fixed $\Delta$, \be 
\la\,\mathcal{O}(t_s)\,\mathcal{O}(0)\,\ra \ \sim \ \lf \frac{1}{\cosh^2\inp{\CJ_0 t_{s}}} \rg^{\Delta}
\ee
is well-behaved and non-trivial. It is neither infinitely rapidly or infinitely slowly varying with respect to $t_s$. If however we write it as a function of cosmic time ($t_{c} = t_s/p$) we will find the cord correlation function to be varying infinitely rapidly in the limit $p\to \infty$:
\be 
\la\,\mathcal{O}(t_c)\,\mathcal{O}(0)\,\ra \ \sim \ \lf \frac{1}{\cosh^2\inp{p\CJ_0 t_{c}}} \rg^{\Delta}
\label{crdcrcosh}
\ee

Now consider the single Fermion two-point function.
Strictly speaking, a single Fermion is not a cord but nevertheless  \eqref{crdcrcosh} gives $G$ by setting $\Delta = 1/p$ (see e.g.  \cite{Maldacena:2016hyu,Lin:2023trc}) giving
\be  
G(t_s) =  \lf \frac{1}{\cosh^2\inp{\CJ_0 t_s}} \rg^{1/p}
\label{Gcosh}
\ee
Since the quantity in the parentheses is less than or equal to $1$, $G(t_s)$ tends to $1$ in the limit $p\to \infty.$ It varies infinitely slowly in string units.

But now consider that same two-point function but with time measured in \textit{cosmic} units:
\be  
G(t_c) =  \lf \frac{1}{\cosh^2\inp{p\CJ_0 t_c}} \rg^{1/p}
\ee
As $p \to \infty$ this function tends to\footnote{Strictly speaking, the techniques that lead to the formula \eqref{Gcosh} do not apply in the regime of times which are order one in cosmic units. Nevertheless, the limiting formula \eqref{Gcos}---which is what we actually need---can be derived using the techniques of \cite{Maldacena:2018lmt}. See also \cite{Roberts:2018mnp} and \cite{Tarnopolsky:2018env}.} 
\be 
G(t_c) \ \sim \  e^{-2\CJ_0 | t_c|}
\label{Gcos}
\ee
The Fermion 2-point function varies infinitely rapidly in string units but in cosmic units it is well-behaved, simply encoding an exponential decay with respect to (we conjecture) the Hawking temperature
\begin{equation}
	\uptau_H = 2\mathcal{J}_0
\end{equation}
This example perfectly illustrates the ``separation of scales": correlators encoding physics at a given scale look perfectly normal in units adapted to that scale but are ill behaved in units adapted to a different, separated, scale. It's interesting to note that had we worked with the Hamiltonian and time in cosmic units from the outset we would have gotten the same answer \eqref{Gcos} for the Fermion 2-point function but without the need to take a limit.

Let's pause here and review. We have found two scales---string and cosmic---from different limits of the same correlation function. In fact the cosmic scale temperature derived from \eqref{Gcos}  is the same as the Tomperature, which for reasons explained in \cite{Lin:2022nss} and Section \ref{Tomp}
is the Hawking temperature.

We saw in section \ref{S:fake} how $T_{\mathrm{cord}}$ and $T_B$ are related. Now we see how SYK produces a new scale which is infinitely separated from the others. Closing the circle requires a bulk explanation of the large separation of $\uptau_{\mathrm{cord}} \sim p\mathcal{J}_0$ and the  Hawking temperature $\uptau_H \sim \mathcal{J}_0$. We will now provide this explaination.

\section{The Blue Shift Factor}
\quad The distinction between the cord temperature and the Hawking temperature was a source of contention between Narovlansky-Verlinde \cite{Narovlansky:2023lfz} and the present authors \cite{Rahman:2023pgt}. The cord and Hawking temperatures are closely related although contrary to the claim of   \cite{Narovlansky:2023lfz} they are not the same;  their numerical values differ by a factor of $p$ which diverges in the double-scaled limit. The large ratio 
\begin{equation}
	\frac{\uptau_{\mathrm{cord}}}{\uptau_{H}} \ \sim \ p \ \to \ \infty
	\label{th<<tomp}
\end{equation}
of
\begin{equation}
	\uptau_{\mathrm{cord}} \ \sim \ p\mathcal{J}_0
\end{equation}
to Hawking temperature
\begin{equation}
	\uptau_H = 2\mathcal{J}_0
\end{equation}
has a simple bulk explanation based on the geometry of de Sitter space. 

The metric of de Sitter space in static patch coordinates has the form
\bea
\mathrm{d}s^2 &=& -f(r)\,\mathrm{d}t^2 +f(r)^{-1}\mathrm{d}r^2 + r^2 \mathrm{d}\Omega_{D-2}^2 \cr \cr
f(r) \eq \lf 1-\frac{r^2}{\ell_{\mathrm{dS}}^2}  \rg
\label{metric}
\eea
The Hawking temperature is the proper temperature experienced by a thermometer at the pode, i.e., at $r=0.$ But generic cords are confined to the stretched horizon which should lie of order  a string length in proper distance from the mathematical horizon at $r = \ell_{\mathrm{dS}}$. The temperature felt by cords is therefore obtained by \it blue-shifting \rm the Hawking temperature by the Blue-Shift Factor (BSF) given by
\be 
\text{BSF} = 
\sqrt{\frac{f(0)}{f(\R)}} = \sqrt{\frac{1}{f(\R)}}
\label{blue}
\ee
i.e. we have that 
\begin{equation}
	\inp{\text{temp seen at stretched horizon}}
	\ = \ \mathrm{BSF}\cdot\inp{\text{temp seen at pode}}
\end{equation}
The notation $f(r_{\mathrm{stretch}})$ denotes the value of $f(r)$ at the stretched horizon $r = r_{\mathrm{stretch}}$.

To calculate the blue shift at the stretched horizon we begin by computing the proper distance from the horizon $\rho(r)$ of a point at radial coordinate $r$ along a slice of constant $t$. This is given by
\be 
\rho(r) = \int_{r}^{\ell_{\mathrm{dS}}} \frac{1}{\sqrt{f(r')}}\,\mathrm{d}r' =   \int \frac{1}{\sqrt{(1-r'^2/{\ell_{\mathrm{dS}}^2})}}\,\mathrm{d}r'.
\ee
Define $r' =\L \cos{\theta}$. The above integral then becomes
\be  
\rho(\theta) = \L \int \mathrm{d}\theta   = \L\,\theta
\ee
or 
\begin{equation}
	\rho(r) = \arccos\inp{r/\ell_{\mathrm{dS}}}
\end{equation}

The BSF  in \eqref{blue}  can be expressed as a function of  $\theta,$
\begin{equation}
	\text{BSF} = \frac{1}{\sqrt{1-r^2/\L^2}} = \frac{1}{\sin{\theta}}.
\end{equation}

Assuming that the stretched horizon lies a proper distance of order the string length from the mathematical horizon at $r = \ell_{\mathrm{dS}}$, i.e. assuming that 
\begin{equation}
	\rho(r_{\mathrm{stretch}}) \ \sim \ \ell_{\mathrm{string}}
\end{equation}
we have that 
\begin{equation}
	\theta_{\mathrm{stretch}} \ \sim \ \frac{\ell_{\mathrm{string}}}{\ell_{\mathrm{dS}}}
\end{equation}
The BSF  \eqref{blue} is therefore given by
\begin{equation}
	\mathrm{BSF} = \frac{1}{\sin\theta} \ \approx \ \frac{\ell_{\mathrm{dS}}}{\ell_{\mathrm{string}}} \ \sim \ p
\end{equation}
(This last equality appeared as equation (4.76) of reference \cite{Rahman:2023pgt}). Thus,
\begin{equation}
	\frac{\uptau_{\mathrm{cord}}}{\uptau_H} \ \sim p
	\label{ratio}
\end{equation}
exactly as required by \eqref{th<<tomp}.

The agreement of \eqref{th<<tomp} (a purely SYK/quantum-mechanical result) and the bulk blueshift calculation is noteworthy. It's not just the agreement itself but also the fact that the bulk blueshift calculation makes use of the de Sitter geometry \eqref{metric} to connect the metric at the pode to the metric at the stretched horizon. It is not sensitive to the detailed behavior of the metric, but it does probe a global property of the static patch, namely the ratio of $f(r)$ at the pode and stretched horizon.

If we now recall that, in the de Sitter region, $T_{\mathrm{cord}}  \sim \CJ_0$, we find from \eqref{ratio} that
\be 
T_H \sim \frac{\CJ_0}{p}.
\ee
This is in opposition to \cite{Narovlansky:2023lfz} where it was assumed that $T_H \sim \CJ_0.$ The lesson is that Hawking temperature is essentially the same thing as the cord temperature but seen from a distance, where it is red shifted by the usual redshift factor  $\sqrt{g_{00}}$.
This accounts for the  hierarchy of temperatures,
\be 
T_B > T_{\mathrm{cord}} >> T_H
\ee
and also provides us with another explanation for the fact that $T_{\mathrm{cord}}$ is order $\mathcal{J}_0$ in string units, which is large enough to melt the generic cords which are trapped in the stretched horizon.

We have now closed the circle and shown how all the temperatures that appear in DSSYK$_{\infty}$ are related to temperatures which appear in the bulk of de Sitter space. The Hawking temperature is nothing but the Tomperature. The cord temperature defined holographically by the periodicity of the cord correlation function (the fake disc temperature) is the local proper temperature at the stretched horizon. That leaves the Boltzmann temperature. As we saw in section 
\ref{corrections} the inverse Boltzmann temperature in string units controls $1/N$ corrections to things like the de Sitter entropy.                 

\section{Summary}

\quad In this final section we will summarize our findings and make some concluding remarks.

\subsection{Input Parameters}
Let's list the parameters that define  DSSYK.
\begin{enumerate}
	\item $N,$ the number of Fermion species. $N$ is the parameter that controls the overall size of the de Sitter space in Planck units\footnote{Here by ``Planck units" we are referring to the strict definition of the Planck length via the Newton constant, $\ell_{\mathrm{Planck}} = G$. We are \textit{not} referring to ``micro units".} (i.e. in units of $G$). It also determines how close the model is to the semiclassical limit\footnote{We use the term semiclassical in the weak sense described in the appendix to \cite{Rahman:2023pgt}. In the weak semiclassical limit large scale gravity becomes classical but matter moving in the de Sitter geometry remains fully quantum.}. The semiclassical limit is simply the large $N$ limit.
	
	\item The SYK locality parameter\footnote{Strictly speaking, in the double scaled limit $p$ is not an independent parameter of the theory, but rather scales as $\sqrt{N}$ due to the finiteness of $\lambda = 2p^2/N$. Here we simply wish to bring attention to the relation between $p$ and the string scale.} $p$ controls the ratio of cosmic and string scale,
	\be  
	\frac{\Ls}{\L} \sim \frac{1}{p}
	\label{ds/string}
	\ee
	The larger is $p$ the smaller the string scale relative to the cosmic scale. In the DSSYK limit the cosmic scale becomes infinite in string units. In other words the theory becomes ``subcosmically local."
	
	\item The parameter $\lambda = 2p^2/N$ is also associated with locality. The ratio of string to micro scale is given by
	\be 
	\frac{\Lm}{\Ls}= \sqrt{\lambda}
	\label{mic/string}
	\ee
	When $\lambda$ is small the string scale becomes large in micro units. $\lambda$ plays the same role in DSSYK$_{\infty}$ as the string coupling constant $g_s$ plays in string theory: when it becomes small the coupling of cords becomes weak.
	
	\item The final input parameter is the Boltzmann temperature which appears in the density matrix of the static patch,
	\be 
	\rho =    \frac{e^{-H/T_B}}{Z}
	\ee
	More precisely, the relevant input parameter will be the value $T_B$ of the Boltzmann temperature in \textit{string units}.
	We'll review its bulk role below.
	
	\end{enumerate}
	
	\subsection{The Temperatures}
	
	\quad Let's now list and review the various temperatures which appeared in our analysis. All of these temperatures are defined in SYK terms but they also have bulk meanings.

	\begin{enumerate}
		
		\item
		The Boltzmann temperature in addition to being a defining parameter controls $1/N$ corrections to the leading $N$ behavior. As an example in Section \ref{corrections}  we worked out the correction to the de Sitter entropy to leading order in $1/T_B.$ For $T_B=\infty$ the entropy is exactly $$\frac{N}{2}\log{2}.$$ The correction is given in \eqref{SfrGR}.  For $\lambda \sim 1$ and $T_B \gtrsim \mathcal{J}_0$ it is of order unity, i.e. a single bit. The interesting thing is that for the overwhelmingly large range of Boltzmann temperatures the density matrix is very close to maximally mixed. In fractional terms the entropy correction is order $1/N.$
		
		\item The tomperature is the change in energy when a single qubit is frozen or
		removed. In an approximate sense that is what happens when a Hawking
		quantum is radiated from the stretched horizon. From the first law $$\Delta E = T \Delta S$$ 
		we see that the Tomperature is approximately the Hawking temperature, i.e., the temperature as seen by an observer at the pode. In cosmic units it is of order unity; in string units it is very small, of order $1/p.$
		
		\item  The cord temperature $T_{\mathrm{cord}}$ is the temperature experienced by generic cords confined to the stretched horizon. In string units it is of order unity which means it is hot enough to ``melt" cords to their constituent Fermions. This explains a number of features of cord correlation functions such as their (leading-order) factorization into products of single Fermion correlators and their overdamped decay.
		
		\quad As an aside we note that the temperatures $T_B$ and $T_{\mathrm{cord}}$ are the temperatures defined by the real and fake discs of \cite{Lin:2023trc}. 
		The relationship between $T_{\mathrm{cord}}$ and $T_B$ was illustrated in figure \ref{myst}. What the figure shows clearly is that for almost the full range of $T_B,$ from  $T_B \gtrsim \mathcal{J}_0$ to $T_B = \infty$ the system is in the de Sitter range. 
		
		\item Finally the Hawking temperature $T_H.$ The Hawking temperature is a bulk concept but it is given by the Tomperature which was defined purely in terms of SYK concepts. It is also given by the thermal decay of single fermion correlators with respect to cosmic time. The ratio of $T_B$ and $T_{\mathrm{cord}}$ to Tomperature/Hawking temperature is order $p$ and diverges in the DSSYK limit. The divergence exactly matches the blue-shift of radiation between the pode and the stretched horizon. This finding bridges the gap between the pode and the stretched horizon and may be
		the most interesting finding of this paper in that it depends on the de Sitter metric in the region between the pode and stretched horizon.
		
	\end{enumerate}

	\subsection{The Transition Region}
	\quad 
	A question that we have not  addressed is: What is the bulk geometry for $T_B < \mathcal{J}_0$? The very small $T_B$ region has been the subject of several papers, see e.g. \cite{Lin:2023trc} and related works. For small $T_B$ and small $\lambda$ the geometry is thought to be close to the usual ``near-AdS$_{2}$" lying between two Schwarzian boundaries. 
	
	The transition region
	$.5 \mathcal{J}_0 \lesssim T_B \lesssim \mathcal{J}_0 $ is something of a mystery: What kind of geometry can interpolate between near-AdS$_2$ for small $T_B$ and de Sitter space for $T_B \gtrsim \mathcal{J}_0$? In approaching this question we are in the land of speculation but we will make our best guess.
	
\begin{figure}[H]
		\begin{center}
			\includegraphics[scale=.33]{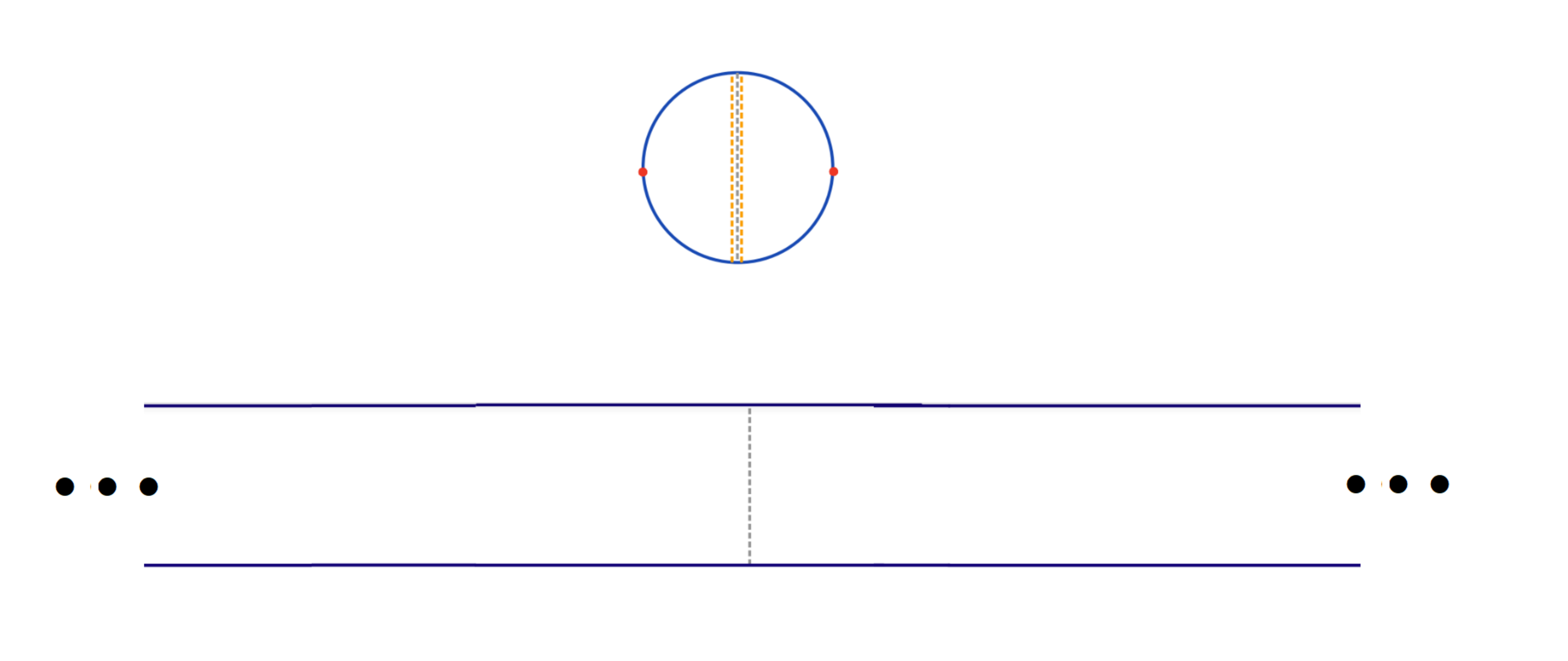}
			\caption{Top: A spatial slice of dS$_3$; Bottom: A spatial slice of NAdS$_2$ equipped with constant dilaton. The dashed grey line is the horizon.}
			\label{dstoads}
		\end{center}
	\end{figure}

	To get some insight into the transition region consider the e.g. $t = 0$ spatial slice of the two limiting situations: $T_B \sim \infty$ and $T_B \sim 0.$ In the first case the geometry is expected to be the dimensional reduction of $(2+1)$ dimensional de Sitter space \cite{Rahman:2022jsf}. The $t = 0$ slice is an interval equipped with a Dilaton\footnote{Here by ``Dilaton" we mean the \textit{total} size of the reduced/transverse dimensions, including any possible large nonfluctuating pieces.} that scales with the static patch radial coordinate. We will represent this slice, accounting for the Dilaton, as a 2D-sphere, i.e. as a slice of dS$_3$ (see the top panel of figure \ref{dstoads}).
	In the second case, the geometry is NAdS$_2$ (``Nearly AdS$_2$", i.e. AdS$_2$ cut off by far-off Schwarzian boundaries) equipped with a nearly-constant Dilaton field. We will represent this slice, accounting for the Dilaton, as a 2D cylinder, as in the bottom of figure \ref{dstoads}.
	
\begin{figure}[H]
		\begin{center}
			\includegraphics[scale=.4]{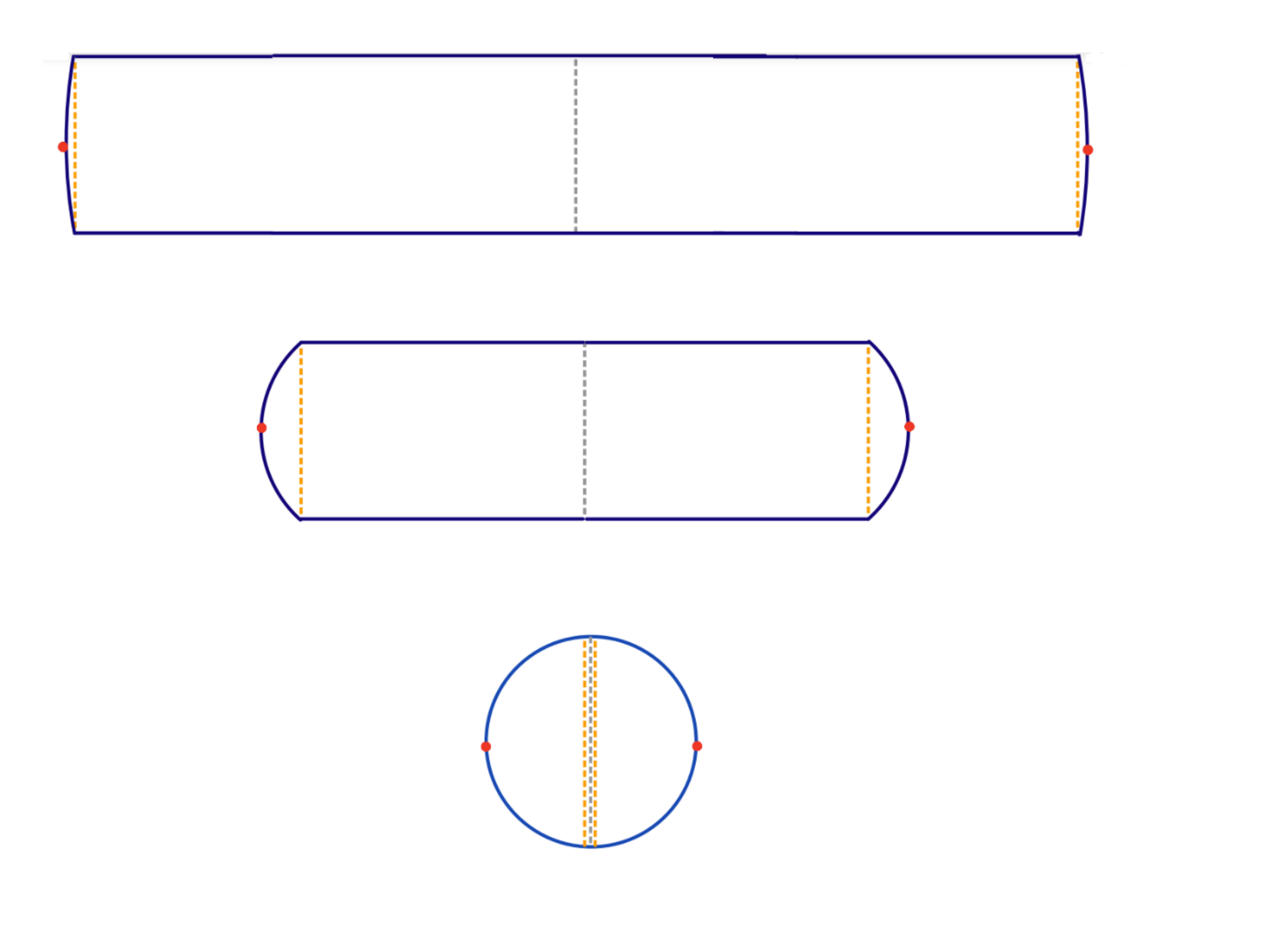}
			\caption{The transition of a spatial slice from low to high temperatures. The dashed grey line is the horizon and the dashed orange lines are the Schwarzian boundaries/domain walls.}
			\label{adstods}
		\end{center}
	\end{figure}

\begin{figure}[H]
		\begin{center}
			\includegraphics[scale=.3]{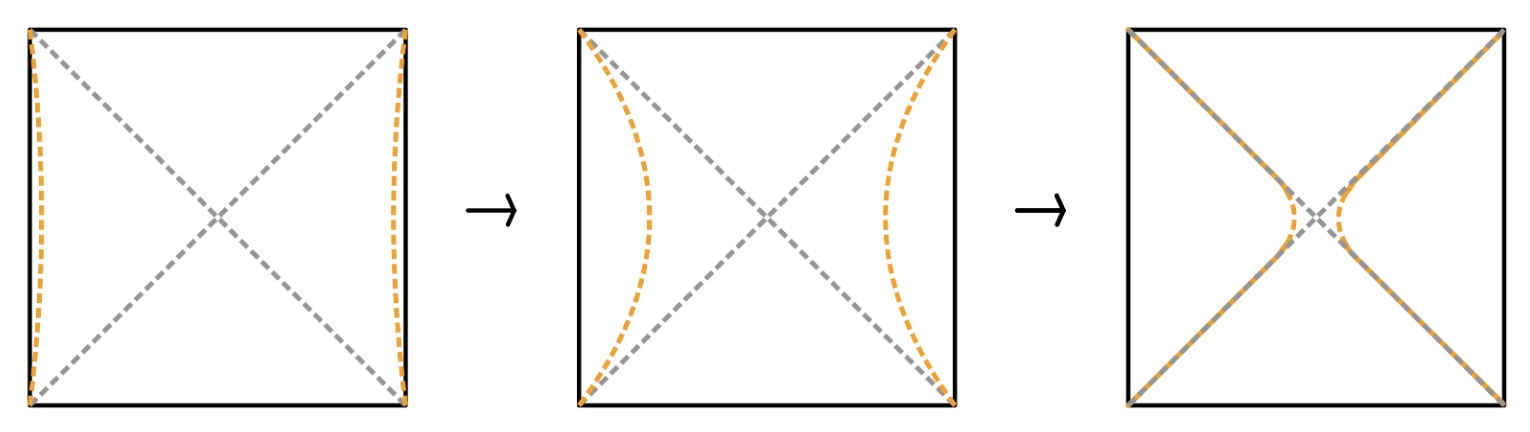}
			\caption{Spacetime (i.e. Penrose diagram) picture of the transition from low (left) to high (right) temperatures.}
			\label{penrose}
		\end{center}
	\end{figure}

	The question is how to smoothly interpolate between these two geometries. We will assume that no sharp transition such as a topology change takes place. Our conjecture  is that  the ends of the AdS geometry are capped off by a de Sitter region whose relative size increases with $T_B$, as shown in figure \ref{adstods}.
	In other words the Schwarzian boundaries of the NAdS$_2$ region---shown as the dashed orange lines---are not ``end of the world" branes, but instead are domain walls separating the negatively curved AdS region from the positively curved de Sitter regions. We propose that as $T_B$ increases the AdS region shrinks until the two Schwarzian boundaries converge and form the stretched horizons of the two-sided de Sitter geometry. We will leave further discussion to a future paper. See figure \ref{penrose} for a spacetime picture of this process.
	
	\section{Conclusions}
	\quad The double-scaled SYK model passes the test of having multiple temperatures, as expected for a holographic theory of de Sitter space. These temperatures---Boltzmann, fake disk/cord, and Tomperature/Hawking---are all defined in terms of the SYK theory with no reference to General Relativity. Nevertheless they all have bulk incarnations. The inverse Boltzmann temperature $\beta_B$ is a measure of $1/N$ corrections to the idealized infinite temperature theory.  The cord temperature $T_{\mathrm{cord}}$ defined by the periodicity of cord correlators is the proper temperature at the stretched horizon. Finally the Tomperature, again defined purely in SYK quantum terms is (up to scaling) the Hawking temperature measured by an observer at the pode. The Tomperature and cord temperature are quantitatively very different but that difference is nothing but the blue shift, predicted by general relativity, relating proper temperatures at the pode and the stretched horizon. Our main conclusion is that the duality between DSSYK$_{\infty}$ and dimensionally reduced dS$_3$ has passed a number of non-trivial tests.
	
	\subsection{Deep Issues} 
	\quad Still, there are unresolved questions raised by the DSSYK$_{\infty}$ duality. One of the most interesting is question of how to formulate ``cord theory."
	As we have explained in a number of places the large $N$ limit of DSSYK$_{\infty}$ exhibits a separation of scales. The intermediate scales centered on the micro and string scales are deep into the flat-space range of parameters. If the duality is correct the theory of singlet cords, far from the horizon in the bulk of the static patch,  must define a relativistic system in flat  spacetime. It should be possible to formulate this cord theory directly in the flat-space limit. The theory might take the shape of
	a discrete light-cone quantization analogous to the BFSS quantization of M-theory, the D0-branes of BFSS being replaced by Majorana Fermions as the carriers of longitudinal momentum.
	
	The existence of DSSYK$_{\infty}$ as an explicit model of de Sitter space runs directly against some of the lore about de Sitter space. For example DSSYK$_{\infty}$ is a completely stable system with a proper ground state and no mechanism  to decay. This is not supposed to be:  all de Sitter vacua are thought to be part of a huge landscape that includes vacua with zero vacuum energy, and can therefore decay---or so the lore says. Being a theory of fluctuations, eternal stable de Sitter space  cannot explain observed universe for the reasons explained in \cite{Dyson:2002pf} but that is a different issue than whether it  is a mathematical
	possibility. DSSYK$_{\infty}$ suggests that it is.
	
	Another bit of lore is that large de Sitter radius requires fine-tuning in order to cancel radiative corrections to the cosmological constant. No fine tuning is required for fomulating \dk-de Sitter,only a large value of $N$ \cite{Banks:2000fe,Fischler}. Can the local matter fields (cords) interact with the gravitational field and ruin the hierarchy of scales? That does not seem possible since it would requre a renormalization of the entropy. But in the $T_B = \infty$ limit the entropy is necessarily exactly  equal to $N\,\frac{\log(2)}{2}$ and no corrections can change that. What's more, even in the finite $T_B$ theory the corrections to the de Sitter entropy are very small.
	
	We leave these deep issues for the future, but point out that  \dk-de Sitter duality is not just a mathematical game. It could have 
	serious implications for our current understanding of cosmology. 
	
\section*{Acknowledgements}
\quad We would like to thank H. Lin for many helpful discussions. A.R. and L.S. are supported in part by NSF Grant PHY-1720397 and by the Stanford Institute of Theoretical Physics.

\end{document}